\documentclass[lettersize,journal]{IEEEtran}
\usepackage{amsmath,amsfonts}
\usepackage{algorithm}
\usepackage{array}
\usepackage[caption=false,font=normalsize,labelfont=sf,textfont=sf]{subfig}
\usepackage{textcomp}
\usepackage{stfloats}
\usepackage{url}
\usepackage{verbatim}
\usepackage{graphicx}
\usepackage{cite}
\usepackage{makecell}

\usepackage{amsthm,amsmath,amssymb}
\usepackage{makecell}
\usepackage{mathrsfs}
  \usepackage{float}
  \usepackage{algorithm}
\usepackage{algpseudocode}
\usepackage{amsmath}
\usepackage{bm}
\usepackage{color}
\usepackage{comment}
\usepackage{booktabs}
\usepackage{leftidx}
\usepackage{epstopdf}
\usepackage{amsmath}
\usepackage{makecell}
\usepackage{multirow}
\usepackage{hyperref}
\usepackage{amssymb}
\usepackage{adjustbox}
\begin{document}

\title{LLM4PG: Adapting Large Language Model for Pathloss Map Generation via Synesthesia of Machines}

\author{Mingran Sun,~\IEEEmembership{Graduate Student Member,~IEEE}, Lu Bai,~\IEEEmembership{Senior Member,~IEEE}, Xiang Cheng,~\IEEEmembership{Fellow,~IEEE}, and Jianjun Wu,~\IEEEmembership{Member,~IEEE}

\thanks{Mingran Sun, Xiang Cheng, and Jianjun Wu are with the State Key Laboratory of Photonics and Communications, School of Electronics, Peking University, Beijing 100871, China (e-mail: mingransun@stu.pku.edu.cn; xiangcheng@pku.edu.cn;
just@pku.edu.cn). 

Lu Bai is with the Joint SDU-NTU Centre for Artificial Intelligence Research (C-FAIR), Shandong University, Jinan 250101, China (e-mail: lubai@sdu.edu.cn).}

}



\maketitle 

\begin{abstract}
In this paper, a novel large language model (LLM)-based pathloss map generation model (LLM4PG) is proposed for sixth-generation (6G) artificial intelligence (AI)-native communication systems via Synesthesia of Machines (SoM). To explore the mapping mechanism between sensing images and pathloss maps, a new synthetic intelligent multi-modal sensing-communication dataset for SoM in uncrewed aerial vehicle (UAV)-to-ground (U2G) scenarios, named SynthSoM-U2G, is constructed, including multiple U2G scenarios with multiple frequency bands and multiple flight altitudes. By adapting the LLM to the cross-modal pathloss map generation for the first time, the proposed LLM4PG introduces a novel framework that enables effective cross-domain alignment between the multi-modal sensing-communication domain and the natural language domain. Furthermore, a task-specific adaptation of the LLM is achieved through fine-tuning, with a properly designed layer selection and activation scheme tailored to the unique demands of cross-modal massive-scale and high-quality pathloss map generation. Compared with the conventional deep learning artificial intelligence generated content (AIGC) models, the proposed LLM4PG enables accurate pathloss map generation and demonstrates strong generalization across various scenarios, frequency bands, and flight altitudes under three-dimensional (3D) high-mobility U2G scenarios. The accuracy and generality of the proposed LLM4PG are validated by comparing simulation results and ray-tracing (RT)-based results. Simulation results demonstrate that the proposed LLM4PG can achieve accurate pathloss map generation with a normalized mean squared error (NMSE) of 0.0454, outperforming the conventional deep learning AIGC model by more than 2.90 dB. The generality of the proposed LLM4PG across different scenarios, frequency bands, and flight altitudes achieves an NMSE of 0.0492, outperforming the conventional deep learning AIGC model by more than 4.52 dB.
\end{abstract}

\begin{IEEEkeywords}
6G communications, large language model (LLM), Synesthesia
of Machines (SoM), pathloss map generation, cross-domain alignment.
\end{IEEEkeywords}

\section{Introduction}
\IEEEPARstart{I}{n} wireless communication systems, accurate characterization of wireless channels is fundamental to core design tasks such as coverage planning and performance optimization \cite{TCoM}. As the cornerstone of communication system design and optimization, channel modeling generally consists of two aspects, namely modeling channel large-scale and small-scale fading characteristics \cite{bai-1}. In particular, large-scale channel modeling focuses on describing the pathloss map of a given scenario, which directly determines link budget, coverage planning, and power control within a certain area, and thus serves as the foundation for ensuring the reliability of communication systems \cite{pathloss}.

For conventional communication system design, the main role of pathloss characterization and modeling is to support link budget calculation and system planning, while also providing a unified comparison platform for algorithm validation. Therefore, it is sufficient to capture the dominant pathloss characteristics without requiring extremely high accuracy. Specifically, three main approaches have been widely adopted for pathloss generation, i.e., stochastic channel modeling, deterministic channel modeling, and the artificial intelligence (AI)-based radio-frequency (RF) data-driven approach. For stochastic channel modeling, standardized channel models represent a typical approach for pathloss generation. In the fifth-generation (5G) communications, Third Generation Partnership Project (3GPP) TR 38.901 channel model \cite{3GPP} was proposed in the frequency band of 0.5-100 GHz, generating pathloss in different typical scenarios. The stochastic channel modeling approach fits empirical formulas from measurement data, enabling low-complexity generation but with limited accuracy. For deterministic channel modeling, ray-tracing (RT)-based methods are widely recognized as typical approaches for site-specific pathloss generation \cite{RT}, which provides high-accuracy point-to-point pathloss, yet suffers from prohibitive computational complexity. For the AI-based RF data-driven approach, the authors in \cite{RF_based-1, RF_based-2} utilized RF data to train a deep neural network, such as a multilayer perceptron (MLP), aiming to improve the pathloss generation accuracy. The aforementioned AI-based RF data-driven approaches in \cite{RF_based-1, RF_based-2} rely on RF information to capture statistical pathloss characteristics more accurately than the stochasitic channel models, but still fall short in producing precise point-to-point pathloss maps. Overall, the aforementioned three main approaches in \cite{3GPP, RT, RF_based-1, RF_based-2} have complementary advantages and drawbacks, and they can sufficiently support conventional system design and algorithm validation. However, in the emerging sixth-generation (6G) communications, communication systems are evolving toward AI-native paradigms, where AI is deeply integrated into system design to significantly enhance performance \cite{NE, AI-native}. Therefore, the scale and quality of data determine the performance ceiling of the 6G AI-native systems, making it essential to generate massive-scale and high-quality pathloss data for algorithm training and optimization. However, among the above three approaches, the stochastic channel modeling approach and the AI-based RF data-driven approach suffer from low accuracy and are thus incapable of generating high-quality pathloss data. Although deterministic channel modeling can achieve high accuracy, its excessive computational complexity makes large-scale data generation impractical. Therefore, these approaches cannot meet the demand of 6G AI-native systems for massive-scale and high-quality pathloss data. This necessitates the development of a more feasible approach that can efficiently generate massive-scale and high-quality pathloss data.

Since the aforementioned three approaches in \cite{3GPP, RT, RF_based-1, RF_based-2} rely solely on RF information to generate pathloss, they lack an understanding and representation of the physical environment, and also suffer from either high computational complexity or low accuracy. To fill this gap, inspired by synesthesia of human, a novel concept, i.e., Synesthesia of Machines (SoM), which refers to intelligent multi-modal sensing-communication integration, was proposed in \cite{SoM}. Within the framework of SoM, the mapping mechanism from physical environment to electromagnetic space can be explored, enabling cross-modal pathloss generation from the physical environment to the electromagnetic space by leveraging easily accessible physical environment information \cite{MMICM}. To leverage the physical environment information, global map-based pathloss generation is introduced, which utilizes satellite map data to support more accurate and flexible pathloss generation. In \cite{map_based-1, map_based-2, map_based-3}, satellite maps were utilized to generate pathloss maps in specific areas based on convolutional neural network (CNN)-based architectures, which enabled pathloss generation in two-dimensional (2D) static ground-to-ground (G2G) communication scenarios. With the evolution of 6G communications, dynamic scenarios are becoming increasingly important and diverse, extending from 2D G2G scenarios with autonomous vehicles to three-dimensional (3D) uncrewed aerial vehicle (UAV)-to-ground (U2G) scenarios \cite{LAE, ped}. In such highly dynamic 6G communication scenarios, pathloss plays an even more critical role, as it directly affects communication link budget, coverage planning, and power allocation for G2G and U2G communication links \cite{U2G}. However, the inherent high-mobility of these scenarios make pathloss generation extremely challenging. The existing global map-based pathloss generation approaches mentioned in \cite{map_based-1, map_based-2, map_based-3} rely on the static global layout provided by global maps, which fails to capture detailed local structures and altitude relevant features, thereby limiting their applicability to pathloss generation in dynamic 6G communication scenarios. To overcome these limitations, we utilized the sensing information from RGB-D images to capture fine-grained environmental details and took the U2G scenario as a representative case, achieving pathloss map generation in dynamic scenarios for the first time \cite{mr_GAN}. However, all the aforementioned AI-based pathloss generation approaches in \cite{RF_based-1, RF_based-2, map_based-1, map_based-2, map_based-3, mr_GAN} is limited to utilizing conventional deep learning models, such as the ResNet and U-Net, for pathloss generation. Constrained by the architectural design and limited parameters, conventional deep learning models struggle to extract and capture the decisive features of complex physical environments and frequency-dependent variations across different scenarios, frequency bands, and flight altitudes, resulting in insufficient accuracy and generalization capability for cross-condition generation \cite{small-model-no}.

Fortunately, with the emergence of large language models (LLMs), unprecedented potential has been demonstrated in multi-task generalization, complex reasoning, and cross-modal understanding \cite{LLM, LLM-ADD}. Compared with conventional deep learning artificial intelligence generated content (AIGC) models \cite{aigc}, LLMs possess stronger reasoning and generalization capabilities, as well as unique advantages in adapting to cross-domain tasks \cite{LLM2}. Recent progress in cross-modal generation, including text, image, and video synthesis, further highlights the versatility of LLMs, and preliminary attempts have also applied LLMs to communication tasks such as channel prediction \cite{LLM4CP}. However, adapting the pre-trained LLM by fine-tuning for cross-modal pathloss map generation is still lacking in the existing literature. By means of task-specific fine-tuning, the LLM can maintain their powerful sequence modeling ability while being efficiently adapted to the pathloss generation task, thus offering a promising solution to fill the gap between existing approaches and the massive-scale and high-quality pathloss data demanded by the 6G AI-native communication systems.

To fill the above gaps, a novel LLM-based pathloss map generation model (LLM4PG) via SoM is proposed for the first time. The proposed general LLM4PG is demonstrated in 3D high-mobility U2G scenarios, utilizing RGB images and depth maps to achieve cross-modal pathloss map generation. Built upon the architecture and methodology of the SynthSoM dataset \cite{SynthSoM}, a new synthetic intelligent multi-modal sensing-communication dataset for SoM in U2G scenarios (SynthSoM-U2G) is constructed to explore the mapping mechanism between sensing images and pathloss maps. By fine-tuning the pretrained LLM to the pathloss map generation, a novel framework introduced in the proposed LLM4PG enables effective cross-modal pathloss map generation. Additionally, the proposed LLM4PG is used as an effective tool to generate massive-scale and high-quality pathloss data for 6G AI-native communication systems. The major contributions and novelties of this paper are summarized as follows.

\begin{enumerate}
\item A novel LLM-based pathloss map generation model via SoM, named LLM4PG, is proposed for the first time, supporting massive-scale and high-quality pathloss map generation for the 6G AI-native communication systems. Across varying scenarios, flight altitudes, and frequency bands, the proposed LLM4PG, powered by the outstanding generalization capability of the LLM, achieves robust and accurate pathloss map generation.

\item A new synthetic intelligent multi-modal sensing-communication dataset for SoM in U2G scenarios, named SynthSoM-U2G, is constructed to explore the mapping mechanism between sensing images and pathloss maps. The constructed SynthSoM-U2G dataset contains 9,490 pathloss maps, 5,660 RGB images, and 5,660 depth maps, covering two U2G scenarios, including urban crossroad and wide lane scenarios, two frequency bands, including 1.6 GHz and 28 GHz, and two flight altitudes, including 50 m and 70~m.

\item By adapting the LLM to the cross-modal pathloss map generation for the first time, the proposed LLM4PG introduces a novel framework that enables effective cross-modal pathloss map generation, which serves as the massive-scale and high-quality data foundation for 6G AI-native communication systems. Specifically, the embedding and decoder architecture is designed to achieve cross-domain alignment among multi-modal sensing-communication domain and the natural language domain, while simultaneously enabling joint embedding of critical factors, including scenarios, flight altitudes, and frequency bands, to support generalization across varying conditions of pathloss map generation. Moreover, a task-specific adaptation of the LLM is achieved through fine-tuning, where an efficient and accurate layer selection and activation scheme is designed, enabling accurate and reliable pathloss map generation across varying conditions.

\item Simulation results demonstrate that the proposed LLM4PG achieves a normalized mean squared error (NMSE) of 0.0454 in full-sample pathloss generation, outperforming the conventional deep learning AIGC model, namely the generative adversarial network (GAN)-based model, by more than 2.90 dB. The generalization performance across different scenarios, frequency bands, and flight altitudes of the proposed LLM4PG achieves an NMSE of 0.0492, outperforming the GAN-based model by more than 4.52 dB. Furthermore, the proposed LLM4PG achieves the full-sample performance of the conventional deep learning AIGC model-based approach using no more than 400 samples in few-shot generalization.

\end{enumerate}

The remainder of this paper is organized as follows. In Section II, the new synthetic intelligent multi-modal sensing-communication dataset for SoM in U2G scenarios, i.e., SynthSoM-U2G, is elaborated. Section III describes the developed LLM4PG, which explores the mapping mechanism between multi-modal sensing images and pathloss maps. In Section IV, the simulation result is presented, and the generalization performance is analyzed and evaluated. Finally, conclusions are presented in Section V.

\section{Dataset Construction}
\begin{figure}[!t]
\centering
\includegraphics[width=0.4\textwidth]{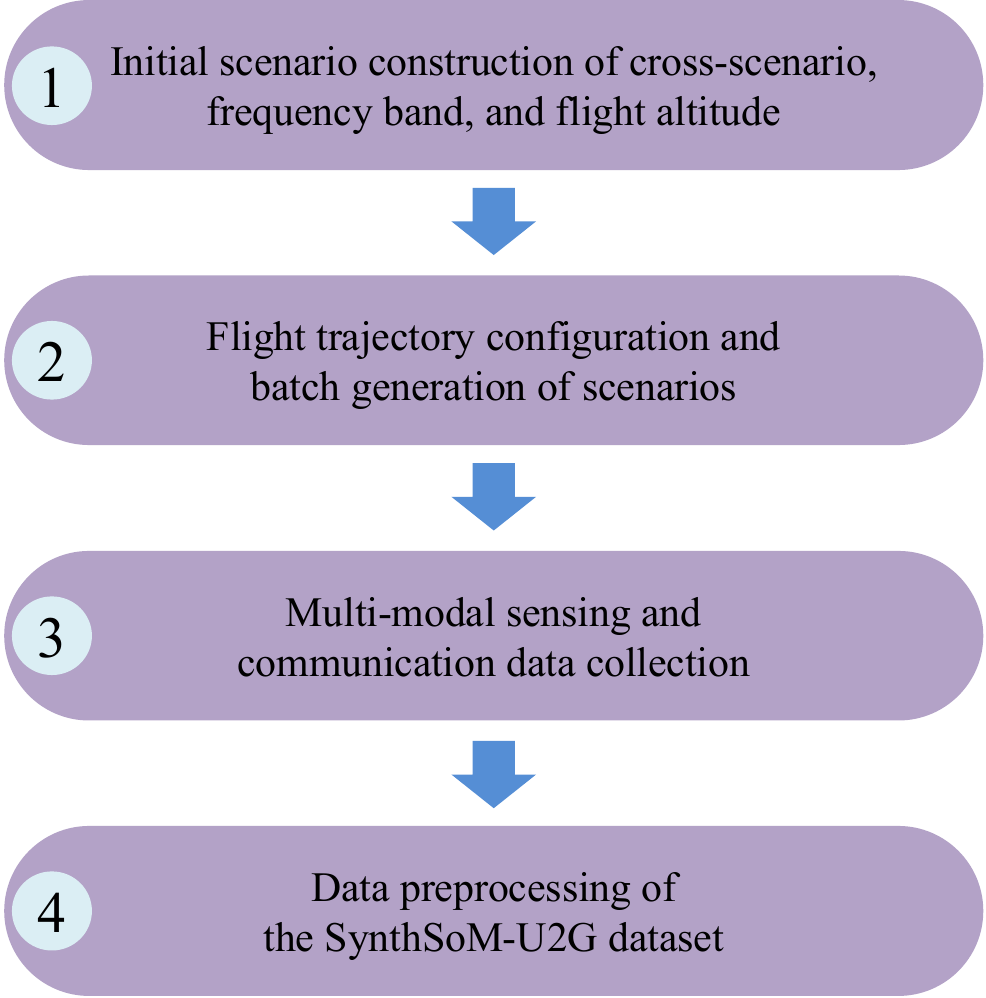}
\caption{Construction process of the SynthSoM-U2G dataset.}
\label{dataset_process}
\end{figure}

Following the framework and methodology established in the SynthSoM dataset \cite{SynthSoM}, a new synthetic intelligent multi-modal sensing-communication dataset for SoM in U2G scenarios, named SynthSoM-U2G, is constructed to explore the mapping mechanism between sensing image and pathloss map. The dataset contains multi-modal sensing images, i.e., RGB images and depth maps, and wireless channel data, i.e., pathloss. Specifically, the constructed dataset consists of 9,490 matched snapshots of RGB images, depth maps, and U2G pathloss maps, including two typical urban scenarios, i.e., urban crossroad and wide lane scenarios, two representative frequency bands of sub-6 GHz and mmWave, i.e., 1.6 GHz and 28 GHz, and two common flight altitudes of UAVs, i.e., 50 m and 70 m. Currently, there is no software that can collect multi-modal sensing data and communication data simultaneously. To fill this gap, we utilize AirSim \cite{AirSim}, which is a simulation plug-in constructed on a 3D Unreal Engine, to collect sensing data and Wireless InSite \cite{Wireless InSite} to collect communication data based on the RT algorithm, achieving precise alignment of sensing data and communication data during the data collection process. Fig.~\ref{dataset_process} illustrates the dataset construction process, including four steps that are described in detail below.

\subsection{Initial Scenario Construction of Cross-Scenario, Frequency Band, and Flight Altitude}

To collect multi-modal sensing and communication data of cross-scenario, frequency band, and flight altitude, the first step is to construct aligned initial scenarios in AirSim and Wireless InSite, respectively. Fig.~\ref{dataset_crossroad} and Fig.~\ref{dataset_widelane} show the urban crossroad scenario and the urban wide lane scenario constructed in AirSim and Wireless InSite. For the initial scenario construction in AirSim, the 3D models of the two typical urban scenarios, i.e., urban crossroad and urban wide lane, are imported into AirSim. Specifically, the 3D model of the urban crossroad scenario is produced by PurePolygons, named Modular Building Set. The roads in the urban wide lane scenario are imported based on data from Beijing Chang'an Avenue on real maps. 

For the initial scenario construction in Wireless InSite, the 3D models of the two scenarios are simplified properly and imported into Wireless InSite, ensuring that the constructed scenario is aligned in both physical environment and electromagnetic space. Furthermore, the flight altitudes of the UAV are set as 50 m and 70 m in the urban crossroad scenario and 200 m in the urban wide lane scenario, which are kept consistent in AirSim and Wireless InSite. Moreover, in Wireless InSite, the surface material of the object model, including buildings and roads, is set to concrete in the two scenarios. Additionally, to enable comprehensive analysis of pathloss generation performance across different frequency bands, the carrier frequencies are set to 28 GHz with 2 GHz communication bandwidth and 1.6 GHz with 20 MHz communication bandwidth, respectively. Therefore, cross-frequency band pathloss data can be collected in both scenarios. Through the above initial scenario construction and parameter configuration, the cross-scenario, frequency band, and flight altitude multi-modal sensing data and pathloss data can be collected in subsequent data acquisition.

\subsection{Flight Trajectory Configuration and Batch Generation of Scenarios}

\begin{figure*}[!t]
\centering
\includegraphics[width=0.90\textwidth]{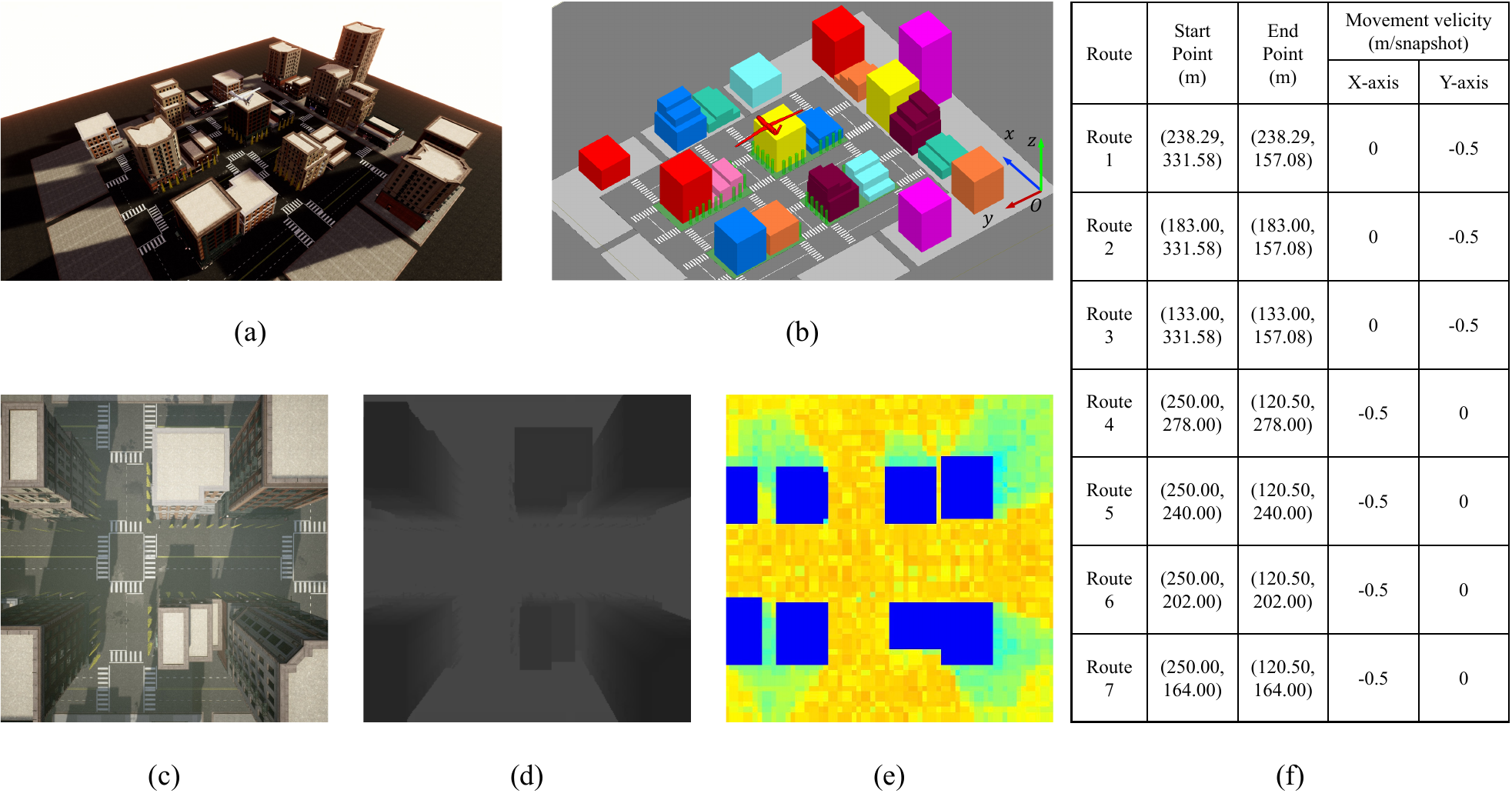}
\caption{The urban crossroad scenario of the constructed SynthSoM-U2G dataset. (a) Scenario demonstration in AirSim. (b) Scenario demonstration in Wireless InSite. (c) An RGB image collected from the scenario at the flight altitude of 50 m. (d) A depth map collected from the scenario at the flight altitude of 50 m. (e) A pathloss map collected from the scenario at the frequency band of 28 GHz. (f) Flight trajectories of the UAV.}
\label{dataset_crossroad}
\end{figure*}

\begin{figure*}[!t]
\centering
\includegraphics[width=0.90\textwidth]{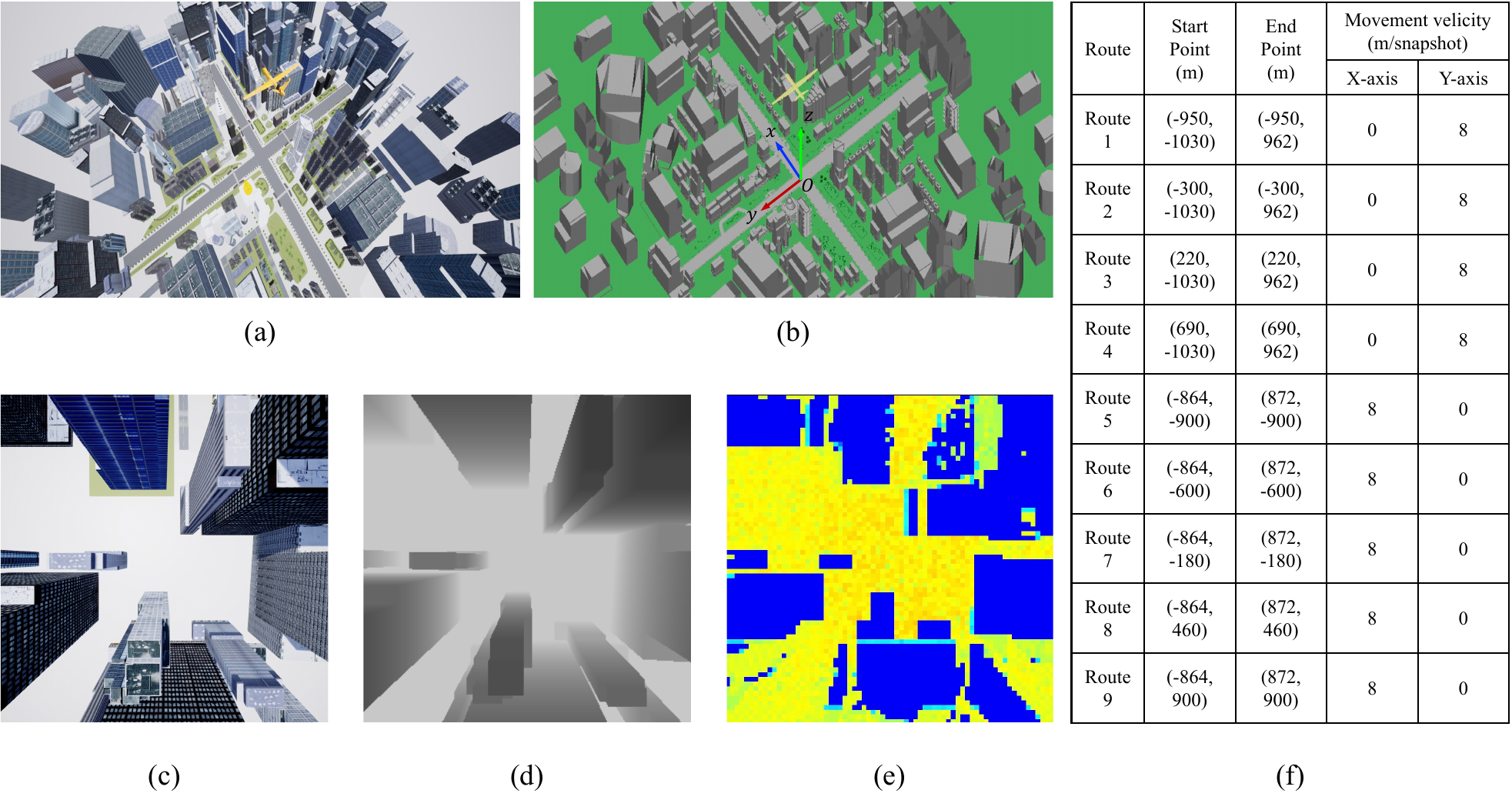}
\caption{The urban wide lane scenario of the constructed SynthSoM-U2G dataset. (a) Scenario demonstration in AirSim. (b) Scenario demonstration in Wireless InSite. (c) An RGB image collected from the scenario at the flight altitude of 200 m. (d) A depth map collected from the scenario at the flight altitude of 200 m. (e) A pathloss map collected from the scenario at the frequency band of 28 GHz. (f) Flight trajectories of the UAV.}
\label{dataset_widelane}
\end{figure*}

To evaluate the generalization ability of the pathloss generation model across diverse flight altitudes, the second step involves configuring flight trajectories and generating scenarios in batches. Based on the constructed initial scenario of cross-scenario, frequency band, and flight altitude in Section II-A, data acquisition in high-mobility scenarios first requires setting the flight trajectory of the UAVs to make them fly according to a predetermined dynamic route, achieving data collection for different snapshots through batch generation of the scenarios. Specifically, for the flight trajectory configuration, the 3D coordinates of the UAVs are set snapshot by snapshot, maintaining strict alignment in both AirSim and Wireless InSite. In Wireless InSite, numerous internal software files are modified through scripts. By batch generation of scenarios at different UAV positions, snapshot-by-snapshot simultaneous dynamic movement of antennas and UAVs is achieved. Note that the UAV flies at a uniform speed along the preset trajectories, where the projections of trajectories at different flight altitudes on the $x$-$y$ plane remain consistent. Fig.~\ref{dataset_crossroad}(a) and Fig.~\ref{dataset_widelane}(a) present the constructed scenarios in AirSim. Fig.~\ref{dataset_crossroad}(b) and Fig.~\ref{dataset_widelane}(b) present the constructed scenarios in Wireless InSite, which are precisely aligned with the scenarios in AirSim. Fig.~\ref{dataset_crossroad}(c)-(d) and Fig.~\ref{dataset_widelane}(c)-(d) illustrate the RGB images and depth maps collected from the urban crossroad scenario and the urban wide lane scenario in AirSim, respectively. Fig.~\ref{dataset_crossroad}(e) and Fig.~\ref{dataset_widelane}(e) respectively illustrate the pathloss maps collected and processed from the urban crossroad scenario and the urban wide lane scenario in Wireless InSite, with the simulation area precisely aligned with the image acquisition range in AirSim. Note that to enhance the diversity of multi-modal sensing data and pathloss data, the trajectories of the UAV are designed to cover a comprehensive range of scenarios, providing a robust data foundation for the LLM4PG and other AI-based models to explore the mapping relationship between physical environment and electromagnetic space. Fig.~\ref{dataset_crossroad}(f) and Fig.~\ref{dataset_widelane}(f) show the trajectories of the UAV in the urban crossroad scenario and the urban wide lane scenario, respectively. 

\subsection{Multi-Modal Sensing and Communication Data Collection}

Efficient collection of multi-modal sensing data and communication data is highly challenging, primarily due to the strict requirements for spatial and temporal alignment between heterogeneous sources. To achieve precisely aligned data collection, the third step is to utilize AirSim and Wireless InSite to collect multi-modal sensing data and pathloss data based on the scenarios batch-generated in Section II-B. For mutli-modal sensing data collection, the RGB-depth (RGB-D) camera is deployed at the bottom of the UAV, enabling the capture of overhead images from the perspective of the UAV. For pathloss data collection, a single antenna, serving as the Tx, is equipped at the same position as the camera at the bottom of the UAV, while an antenna grid is deployed on the ground in the corresponding sensing area as the receiver (Rx), which can be viewed as a distributed multiple-input multiple-output (MIMO) array or a dense set of potential Rx locations, supporting one-shot pathloss generation for dynamic UAV trajectories and enabling applications such as network planning and coverage analysis. Note that the perception range of the RGB-D camera is the same as the coverage range of the Rx antenna grid at the same flight altitude, achieving sensing and communication data collection in the matched area. Specifically, the data collection ranges at the flight altitude of 50 m and 70 m in the urban crossroad scenario are 99${\times}$99 ${\rm{m}^2}$ and 143${\times}$143 ${\rm{m}^2}$, respectively. The data collection range at the flight altitude of 200 m in the urban wide lane scenario is 473${\times}$473 ${\rm{m}^2}$. To accommodate pathloss data collection at different flight altitudes, the number of antennas in the Rx grid at the flight altitudes of 50 m and 70 m in the urban crossroad scenario is 30${\times}$30 and 50${\times}$50, respectively. The number of antennas in the Rx grid in the urban wide lane scenario is 80${\times}$80. 

Under the above settings, for sensing data collection, the RGB-D images, including RGB images and depth maps, are automatically acquired and saved in AirSim based on the predefined flight trajectory and parameter configurations. For pathloss data collection, the scenarios batch-generated in Wireless InSite are queued for simulation through scripts. Specifically, for multi-modal sensing data collection, 1,830 snapshots of RGB images and 1,830 snapshots of depth maps are collected in the urban crossroad scenario at the flight altitude of 50 m, and the same applies at the flight altitude of 70 m. 2,000 snapshots of RGB images and 2,000 snapshots of depth maps are collected in the urban wide lane scenario at the flight altitude of 200 m. For pathloss data collection, in the urban crossroad scenario at the flight altitude of 50 m, a total of 1.647 M pairs of pathloss values of U2G communication links are collected, including data under the 28 GHz frequency bands. In the urban crossroad scenario at the flight altitude of 70 m, a total of 9.15 M pairs of pathloss values of U2G communication links are collected, including data under the 1.6 GHz and 28 GHz frequency bands. In the urban wide lane scenario at the flight altitude of 200 m, a total of 25.6 M pairs of pathloss values of U2G communication links are collected, including data under the 1.6 GHz and 28 GHz frequency bands. 

\subsection{Data Preprocessing of the SynthSoM-U2G Dataset}

The preprocessing of raw data collected from Section II-C plays a critical role in shaping the performance of the proposed LLM4PG, as it directly influences data quality, cross-modal alignment, and representation consistency. In this subsection, the fourth step, i.e., data preprocessing of the constructed SynthSoM-U2G dataset, is clarified. For multi-modal sensing data preprocessing, RGB images and depth maps collected under the certain scenario and at the certain flight altitude are matched to form single-snapshot RGB-D data directly. For pathloss data preprocessing, the raw data collected from Wireless InSite contains point-to-point pathloss values of U2G communication links. The pathloss values, which range from 0 to 255 dB, are directly assigned to the pixel values of the pathloss map, which also range from 0 to 255. In total, 9,490 snapshots of matched RGB images, depth maps, and pathloss maps of cross-scenario, frequency band, and flight altitude are acquired. Table~\ref{dataset} summarizes the SynthSoM-U2G dataset, including the quantity of multi-modal sensing data and pathloss data across scenarios, frequency bands, and flight altitudes.

\begin{table*}[!t]
\renewcommand\arraystretch{1.25}
\caption{Key Parameters Utilized in the Simulation.\label{tab:table1}}
\centering
\begin{tabular}{|c|c|c|c|c|c|}
\hline
\textbf{Scenario} & \textbf{Flight altitude} & \textbf{Frequency band} & \textbf{Number of pathloss maps} &\textbf{Number of RGB images} & \textbf{Number of depth maps} \\ 
\hline
\multicolumn{1}{|c|}{\multirow{3}{*}{\makecell[c]{Urban crossroad}}} & \multicolumn{1}{c|}{\multirow{1}{*}{50 m}} & \multicolumn{1}{c|}{\multirow{1}{*}{\makecell[c]{28 GHz}}} & \multicolumn{1}{c|}{\multirow{1}{*}{\makecell[c]{1,830}}} & \multicolumn{1}{c|}{\multirow{1}{*}{\makecell[c]{1,830}}}& \multicolumn{1}{c|}{\multirow{1}{*}{\makecell[c]{1,830}}} \\ 

    \cline{2-6}
    \multicolumn{1}{|c|}{}& \multicolumn{1}{c|}{\multirow{2}{*}{\makecell[c]{70 m}}} & \multicolumn{1}{c|}{\multirow{1}{*}{\makecell[c]{28 GHz}}} &  \multicolumn{1}{c|}{\multirow{1}{*}{\makecell[c]{1,830}}} & \multicolumn{1}{c|}{\multirow{2}{*}{\makecell[c]{1,830}}}& \multicolumn{1}{c|}{\multirow{2}{*}{\makecell[c]{1,830}}} \\
    \cline{3-4} 
    {} & {} & 1.6 GHz & 1,830 & {} & {}\\
\hline
\multicolumn{1}{|c|}{\multirow{2}{*}{\makecell[c]{Urban wide lane}}} & \multicolumn{1}{c|}{\multirow{2}{*}{200 m}} & \multicolumn{1}{c|}{\multirow{1}{*}{\makecell[c]{28 GHz}}} & \multicolumn{1}{c|}{\multirow{1}{*}{\makecell[c]{2,000}}} & \multicolumn{1}{c|}{\multirow{2}{*}{\makecell[c]{2,000}}}& \multicolumn{1}{c|}{\multirow{2}{*}{\makecell[c]{2,000}}} \\ 
    \cline{3-4} 
    {} & {} & 1.6 GHz & 2,000 & {} & {} \\
    
\hline

\end{tabular}
\label{dataset}
\end{table*}

\section{Framework of the Proposed LLM4PG for Pathloss Map Generation}

\begin{figure*}[!t]
\centering
\includegraphics[width=0.95\textwidth]{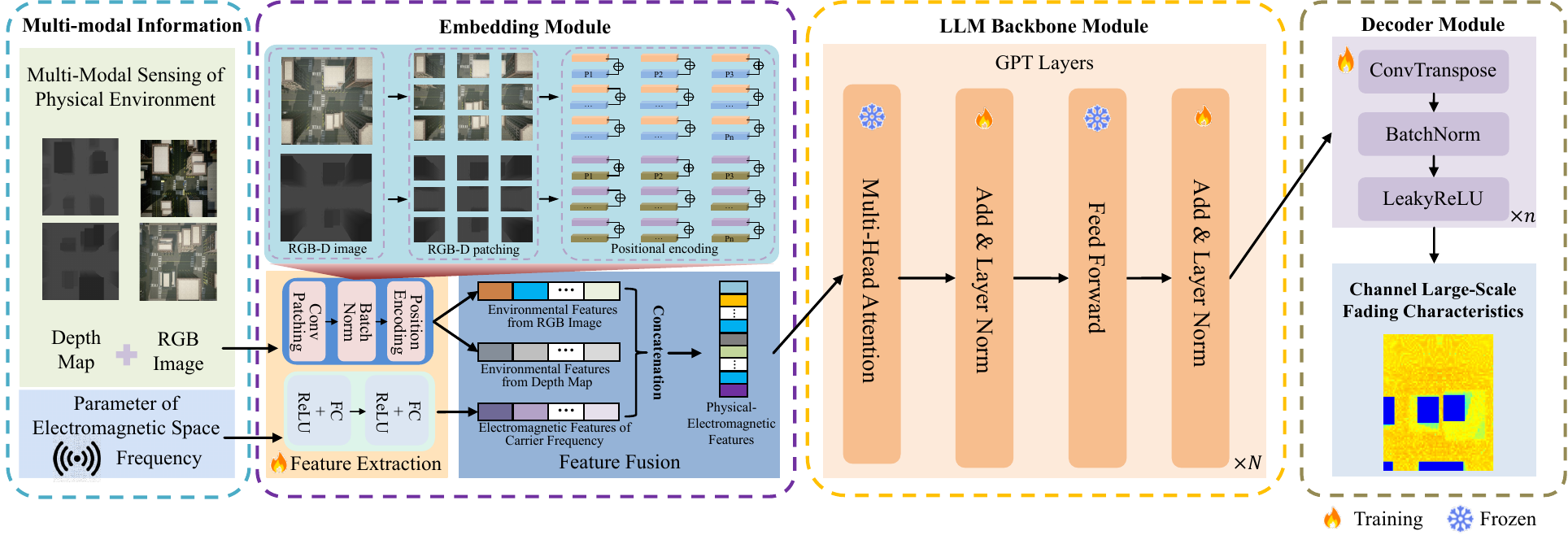}
\caption{An illustration of the network architecture of the proposed LLM4PG.}
\label{network}
\end{figure*}

To generate massive-scale and high-quality pathloss map data for the 6G AI-native communication systems, a novel LLM-based pathloss map generation model via SoM, named LLM4PG, is proposed for the first time. The proposed LLM4PG introduces a novel framework that enables effective cross-domain alignment of multi-modal sensing-communication domain with the natural language domain. Specifically, the embedding and decoder modules are properly designed to achieve cross-domain alignment between the multi-modal sensing-communication domain and the natural language domain, while jointly embedding critical factors, including scenarios, flight altitudes, and frequency bands, to support robust generalization under diverse conditions. Furthermore, a task-specific adaptation of the GPT-2 \cite{GPT-2} is performed through fine-tuning, where an efficient and accurate layer selection and activation scheme is developed to optimize the ability of the proposed LLM4PG in generating accurate pathloss maps. The overall architecture of LLM4PG consists of three modules, including the embedding module, the LLM backbone module, and the decoder module, as illustrated in Fig.~\ref{network}. The design details of each module and the training process are described in detail below.

\subsection{Embedding Module}
To enable LLM adaptation of multi-modal sensing data, including RGB images, depth maps, and communication frequency information, a tailored patching and positional encoding strategy is proposed. Unlike the standard Vision Transformer (ViT) \cite{ViT}, which processes only visual data, the proposed strategh performs early-stage feature fusion across sensing and communication modalities and encodes spatial-frequency semantics into a unified sequential representation. The fusion-based embedding strategy facilitates effective cross-domain alignment between the multi-modal sensing-communication domain and the natural language domain, while also enhancing generalization across varying scenarios, flight altitudes, and frequency bands, thereby enabling the generation of massive-scale and high-quality pathloss data that serves as a critical data foundation for 6G AI-native communication systems. The embedding module can be divided into two components, i.e., feature extraction and feature fusion. For the feature extraction, RGB images and depth maps in physical environment and carrier frequency information in electromagnetic space are considered and processed. Specifically, for RGB images and depth maps, feature extraction is performed using a combination of image patching and positional encoding, while for frequency information, a multilayer perceptron (MLP) is employed for feature extraction. Based on the feature extraction, the environmental features from RGB images and depth maps, and the electromagnetic features of carrier frequency are extracted from the multi-modal information from physical environment and electromagnetic space. Furthermore, for the feature fusion, three types of features, including environmental features extracted from RGB images, environmental features extracted from depth maps, and electromagnetic features of carrier frequency, are concatenated in the feature dimension to obtain physical-electromagnetic features that simultaneously characterize both the physical environment sensing information and the electromagnetic space propagation characteristics. Finally, physical-electromagnetic features suitable for adapting to LLMs are obtained, which can be embedded into the natural language domain. The specific model details of the embedding module, including feature extraction and feature fusion, are elaborated in detail below.

In the feature extraction component, the RGB images and depth maps of physical environment and the carrier frequency parameter of electromagnetic space are considered in the network design. The Transformer-based LLM model only takes a one-dimensional (1D) sequence of token embeddings as input. However, RGB images and depth maps from the physical environment are two-dimensional data. Therefore, when extracting RGB-D image features, it is necessary to convert the two-dimensional (2D) RGB-D images into a sequence of token embeddings. To enable LLM adaptation of multi-modal sensing image data, the patching and positional encoding strategy is adopted to tokenize spatial semantics into a sequential representation. It naturally converts 2D RGB-D images into a 1D sequence of token embeddings compatible with Transformer-based LLMs, enabling seamless integration with electromagnetic frequency features. Furthermore, each patch can focus on local feature regions, improving feature extraction efficiency. Subsequently, the interactions between patches are captured by the self-attention mechanism of Transformers, thereby enhancing the understanding of the global physical environment. Positional encoding explicitly adds positional information, enabling the Transformer to recognize the original spatial layout of each patch in the RGB-D images, thereby preserving the spatial structure. To be specific, the data matrix of RGB images and depth maps 
$\mathbf{M} \in \mathbb{R}^{r_\text{x} \times r_\text{y} \times H}$ are converted into a feature map matrix $\mathbf{\tilde{M}}_\text{p}$, including a sequence of patches, which can be expressed as 

\begin{equation}
\label{deqn_ex1}
\mathbf{\tilde{M}}_\text{p} = \mathcal{B}(\mathcal{C}(\mathbf{M})) \in \mathbb{R}^{E \times n_\text{p} \times n_\text{p}}
\end{equation}
where $r_\text{x} \times r_\text{y}$ is the resolution of the original image, $H$ is the number of channels of the input image. $\mathcal{C}(\cdot)$ and $\mathcal{B}(\cdot)$ represent the convolutional patching layer and the batch normalization operation, respectively. The batch normalization operation accelerates neural network training and enhances model stability by standardizing each batch of data. $E$ is the dimension of the constant latent vector of the Transformer in the LLM backbone. $n=n_\text{p}^2$ represents the number of patches into which an image is divided, which can be calculated as 
\begin{equation}
\label{deqn_ex1}
n = \frac{r_\text{x} \times r_\text{y}}{k^2}
\end{equation}
where $k$ represents the kernel size of the convolutional layer, which is equal to the number of strides. Moreover, to enhance the network's ability to capture nonlinear feature representations, a rectified linear unit (ReLU) operation is applied to the obtained feature map $\mathbf{\tilde{M}}_\text{p}$ as  
\begin{equation}
\label{deqn_ex1}
\mathbf{M}_\text{p}^{'} = \mathcal{R}(\mathbf{\tilde{M}}_\text{p})
\end{equation}
where $\mathcal{R}(\cdot)$ represents the ReLU operation, which can be expressed as 
\begin{equation}
\label{deqn_ex1}
\mathcal{R}(x) = \max{(0,x)}
\end{equation}
where $x\in \mathbb{R} $ can represent any numerical input. Furthermore, to adapt to the natural language domain of the LLM, the feature map is further flattened and transposed, which can be expressed as 
\begin{equation}
\label{deqn_ex1}
\mathbf{M}_\text{p} = \mathcal{F}^\text{T}(\mathbf{M}_\text{p}^{'})
\end{equation}
where $\mathcal{F}^\text{T}(\cdot)$ represents the flattening and transposition operation, which converts $\mathbf{M}_\text{p}^{'} \in \mathbb{R}^{E \times n_\text{p} \times n_\text{p}} $ into $\mathbf{M}_\text{p} \in \mathbb{R}^{n \times E}$. 

To preserve the spatial positional structure among patches in the original images, the positional encoding is further utilized to process the obtained feature $\mathbf{M}_\text{p}$. Specifically, the random positional encoding is utilized in the embedding module. Random positional encoding has unique advantages over fixed positional encoding, such as sine or cosine encoding, or other dynamic positional encoding approaches, such as relative positional encoding. By randomly generating position identifiers, the random positional encoding approach provides greater flexibility and generalization, adapting to different sequence lengths and task requirements while avoiding excessive reliance on specific positional information, thereby enhancing robustness. The random positional encoding is noted as $\mathbf{M}_\text{e} \in \mathbb{R}^{n \times E}$. Then, the positional encoding is added to the obtained feature $\mathbf{M}_\text{p}$, which can be expressed as 
\begin{equation}
\label{deqn_ex1}
\mathbf{M}_\text{pe} =  \mathbf{M}_\text{p} + \mathbf{M}_\text{e}
\end{equation}
where $\mathbf{M}_\text{pe}$ represents the environmental features from RGB images and depth maps, which can be expressed as $\mathbf{M}_\text{pe}^\text{RGB} \in \mathbb{R}^{n_r \times E}$ and $\mathbf{M}_\text{pe}^\text{dep} \in \mathbb{R}^{n_d \times E}$, respectively. $n_r$ represents the number of patches into which the RGB image is divided. $n_d$ represents the number of patches into which the depth map is divided. The difference between obtaining $\mathbf{M}_\text{pe}^\text{RGB}$ and $\mathbf{M}_\text{pe}^\text{dep}$ lies in that, for RGB images, $H=3$, while for depth maps, $H=1$.

For electromagnetic feature extraction, an MLP is utilized to process the carrier frequency parameter in numerical form. Specifically, the frequency $f_c$ is processed through an MLP to obtain the electromagnetic feature $M_\text{f} \in \mathbb{R}^{1 \times E}$, which is expressed as 
\begin{equation}
\label{deqn_ex1}
\mathbf{M}_\text{f} =  \mathcal{P}(f_c)
\end{equation}
where $\mathcal{P}(\cdot)$ represents the MLP network. Overall, the environmental feature $\mathbf{M}_\text{pe}^{\text{RGB}}$ extracted from the RGB images, the environmental feature $\mathbf{M}_\text{pe}^{\text{dep}}$, and the electromagnetic feature $\mathbf{M}_\text{f}$ are obtained based on the feature extraction component. 

In the feature fusion component, to achieve the feature fusion of the physical environment and electromagnetic space, the three features extracted from multi-modal sensing-communication information are concatenated in the patch dimension while ensuring consistency in the embedding dimension. Specifically, the physical-electromagnetic feature $\mathbf{M}^{\text{PE}}$ is calculated as 
\begin{equation}
\label{deqn_ex1}
\mathbf{M}^{\text{PE}} = \mathcal{C}(\mathbf{M}_\text{pe}^{\text{RGB}},\mathbf{M}_\text{pe}^{\text{dep}},\mathbf{M}_\text{f}) \in \mathbb{R}^{(n_r+n_d+1) \times E}
\end{equation}
where $\mathcal{C}(\cdot,\cdot,...,\cdot)$ represents the concatenation of features in the patch dimension while maintaining consistency in the embedding dimension. Overall, based on the embedding module, the physical-electromagnetic feature $\mathbf{M}^{\text{PE}}$ is obtained, achieving the embedding from the multi-modal information domain to the natural language domain.

\subsection{LLM Backbone Module}

To explore the mapping relationship between physical-electromagnetic features and pathloss representations in the electromagnetic space, an LLM is fine-tuned through a task-specific adaptation strategy that includes the design of an efficient and accurate layer selection and activation scheme. This adaptation enables the LLM to effectively transfer its pre-trained general knowledge and sequence modeling capabilities to the pathloss generation task, resulting in accurate and generalizable pathloss prediction across diverse scenarios, flight altitudes, and frequency bands, which in turn facilitates the rapid generation of massive-scale and high-quality pathloss map generation essential for training and evolving 6G AI-native communication systems.

To fine-tune an LLM to explore the mapping relationship between physical environments and electromagnetic spaces, GPT-2 is selected as the backbone module due to its robust architecture and suitability for complex feature integration tasks. As a Transformer-based generative LLM, GPT-2 excels in sequence modeling, enabling effective processing of multi-modal feature data, such as physical-electromagnetic features concatenated along the patch dimension and embedded into the natural language domain. Compared to conventional deep learning models, GPT-2 offers several distinct advantages in exploring the mapping relationship between physical environment information and pathloss in electromagnetic space. Firstly, its unidirectional autoregressive structure enables efficient capture of temporal–spatial dependencies in multi-modal sensing-communication and pathloss sequences, enhancing representational capacity for the complex mapping relationship exploration tasks. Secondly, the moderate model size of GPT-2 balances computational cost and generation capability, requiring fewer resources for fine-tuning while outperforming conventional deep learning models in generalization. Specifically, the Transformer block is configured with 6 layers, where the weights of the multi-head attention and feed-forward modules in each layer are frozen, and only the remaining components are activated and fine-tuned. Finally, the semantic knowledge embedded in its pre-trained weights supports more effective decoding of complex environmental semantics from sensing image inputs, which directly benefits cross-modal pathloss generation.

The physical-electromagnetic features $\mathbf{M}^{\text{PE}}$ obtained from the embedding module are mapped by the LLM backbone module to the electromagnetic feature space of pathloss, acquiring electromagnetic features that characterize pathloss. Specifically, the electromagnetic features of pathloss data are expressed as $\mathbf{M}^\text{L}$, which is given by
\begin{equation}
\label{deqn_ex1}
\mathbf{M}^\text{L} = \mathcal{B}_\text{LLM}(\mathbf{M}^{\text{PE}})
\end{equation}
where $\mathcal{B}_\text{LLM}(\cdot)$ represents the LLM backbone network, including the first $n_\text{L}$ layers of the pre-trained GPT-2 model.

\subsection{Decoder Module}
The decoder module decodes the electromagnetic space features into the pathloss map. First, to reconstruct high-resolution pathloss images from compressed electromagnetic features, the transposed convolution layer is utilized due to its effectiveness for upsampling latent representations. Second, to mitigate issues like vanishing gradients, the batch normalization operation is utilized to stabilize and accelerate training by normalizing intermediate feature distributions. Third, the LeakyReLU \cite{LeakyReLU} operation is utilized to introduce non-linearity while preventing the ``dying ReLU" problem by allowing a small gradient for negative inputs, thus preserving information flow and enhancing feature expressiveness. Finally, by integrating and concatenating the three components, including the transposed convolution layer, the batch normalization operation, and the LeakyReLU operation, the upsampling and decoding of electromagnetic features are achieved, thereby enabling the generation of pathloss maps. The process of the pathloss map generation in the decoder module can be expressed as 
\begin{equation}
\label{deqn_ex1}
\mathbf{M}^\text{PL} = (\mathcal{L_R} \cdot \mathcal{B} \cdot \mathcal{C}^\text{T})^{n_\text{D}} (\mathbf{M}^{\text{L}}) \in \mathbb{R}^{p_\text{x} \times p_\text{y}}
\end{equation}
where $\mathcal{C}^\text{T}(\cdot)$ represents the transposed convolution layer. $\mathcal{B}(\cdot) $ represents the batch normalization operation. $\mathcal{L_R}( \cdot )$ represents the LeakyReLU operation. $n_\text{D}$ represents the number of concatenated submodules, which means that the $n_\text{D}$ submodules, each composed of components the transposed convolution layer, the batch normalization operation, and the LeakyReLU operation, are concatenated in series to form the decoder module. $p_\text{x} \times p_\text{y}$ represents the resolution of the generated pathloss map.

In summary, the proposed LLM4PG model addresses the challenges of heterogeneous data representation and cross-domain alignment in cross-modal pathloss generation by adapting a pre-trained LLM in the 6G dynamic U2G communication scenarios for the first time. Leveraging the generalization and sequence modeling capabilities of the GPT-2 backbone, LLM4PG establishes a novel framework for cross-modal pathloss map generation that inherently serves as a tool to generate massive-scale and high-qualilty pathloss map data, forming the critical data foundation for the 6G AI-native communication systems. Specifically, the embedding and decoder modules are designed to achieve effective cross-domain alignment between the multi-modal sensing-communication domain and the natural language domain, while jointly embedding physical environmental and communication factors, including scenarios, flight altitudes, and frequency bands, to support robust generalization across varying conditions. Moreover, a task-specific adaptation is performed through fine-tuning, where a lightweight yet effective layer selection and activation scheme is developed to enhance the LLM4PG accuracy and reliability in generating pathloss maps under diverse conditions.

\section{Simulation Results and Analysis}

In this section, the simulation configurations are shown. Furthermore, the full-sample performance of the proposed LLM4PG in the two typical urban scenarios is evaluated. Finally, the few-shot generalization performance across different scenarios, different frequency bands, and different flight altitudes is analyzed. 

\subsection{Setup}
For the detailed hyper-parameters of the LLM4PG network design and training, the hyper-parameters are listed in Table~\ref{hp}. For the LLM4PG network design, the pre-trained GPT-2, which has a feature dimension of $E = 768$, is utilized to explore the mapping relationship between features in physical environment and electromagnetic space. To balance the trade-off between complexity and performance, the first $n_\text{L}=6$ layers of the pre-trained GPT-2 are utilized for fine-tuning. Note that in the pre-trained GPT-2 model, the self-attention and feed-forward layers are frozen, with the remaining parameters open for training. Since these fixed layers account for most of the model's parameters, the portion available for training is comparatively small. For the LLM4PG network training, the PyTorch framework is utilized with the optimizer of Adaptive Moment Estimation (ADAM) \cite{ADAM}. During the training process, mean squared error (MSE) loss is utilized to optimize the network parameters.  The dataset is divided into the training set, validation set, and test set, in the proportion of 3:1:1. The network is trained on the training set, evaluated on the validation set to adjust hyper-parameters, and finally tested for performance on the test set.

\begin{table}[h!]
\renewcommand\arraystretch{1.5}
\centering
\caption{Hyper-Parameter for LLM4PG Design and Training}
\begin{tabular}{|c|c|}
\hline
\textbf{Parameter} & \textbf{Value} \\ \hline
Size of convolution kernels in embedding module@$k$ & 8 \\ \hline
The number of patches into which an image is divided@$n$ & 64 \\ \hline
The number of submodules in the decoder module@$n_\text{D}$ & 3 \\ \hline
GPT-2 feature dimension@$E$ & 768 \\ \hline
The number of GPT-2 layers@$n_\text{L}$ & 6 \\ \hline
Batch size & 128 \\ \hline
Learning rate & $1 \times 10^{-4}$ \\ \hline
Epochs & 200 \\ \hline
Optimizer & ADAM \\ \hline
Loss function & MSELoss \\ \hline

\end{tabular}
\label{hp}
\end{table}

For the baseline of the pathloss generation performance, a conventional deep learning AIGC model, which is a GAN-based model, is utilized to generate pathloss maps from the RGB-D images. The generator of the GAN-based model is composed of a ResNet network. The discriminator is composed of a CNN network. The pathloss generation results from the GAN-based model are utilized to provide a comparative reference for the performance improvement of the proposed LLM4PG. 

\subsection{Full-Sample Pathloss Map Generation Performance of LLM4PG}

\begin{figure}[!t]
\centering
\includegraphics[width=0.40\textwidth]{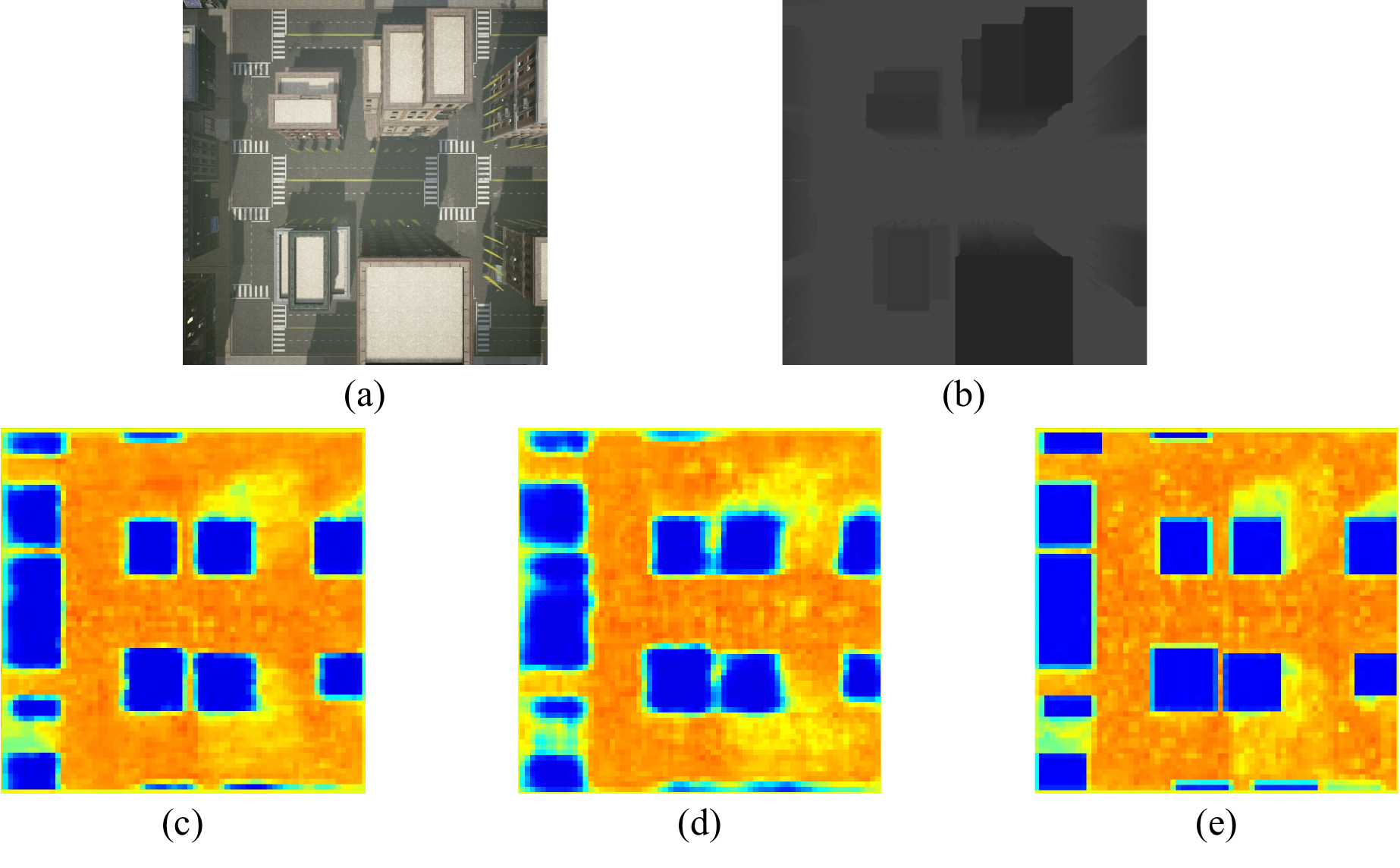}
\caption{Pathloss generation results of LLM4PG and GAN-based model baseline. (a) The input RGB image for the pathloss generation model. (b) The input depth map for the pathloss generation model. (c) The pathloss generation result of the \textbf{LLM4PG}. (d) The pathloss generation result of the \textbf{GAN-based model}. (e) The \textbf{ground truth} of the pathloss map simulated by RT.}
\label{fs-generation}
\end{figure}

The full-sample generation performance of the LLM4PG under different conditions is elaborated in this section. Specifically, the average normalized mean squared error (NMSE) is utilized to evaluate the generation performance, which can be calculated as 
\begin{equation}
\label{deqn_ex1}
\text{NMSE} = \mathbb{E} \left\{ 
\frac{
\sum_{i=1}^{N} \left\| \hat{\mathbf{P}} - \mathbf{P} \right\|^2
}{
\sum_{i=1}^{N} \left\| \mathbf{P} \right\|^2
}
\right\}
\end{equation}
where $\mathbf{P}$ represents the ground truth of pathloss generation simulated by RT, $\hat{\mathbf{P}}$ represents the pathloss generation result by the proposed LLM4PG or the GAN-based small model. $\mathbb{E}[\cdot]$ represents the statistical expectation computed over all samples in the test set. $i$ denotes the index of the $i$-th sample among all $N$ samples in the test set. $\left\| \cdot \right\|^2$ denotes the squared Frobenius norm, which is commonly used to measure the difference between two matrices in terms of element-wise squared error. Table~\ref{full-samples} presents the generation performance across different scenarios, frequency bands, and flight altitudes under full-sample training, i.e., the test results of the LLM4PG trained on the entire training dataset. The impact of different input modalities of sensing image on pathloss generation performance is compared in the urban crossroad scenario at the flight altitude of 50 m and the carrier frequency of 28 GHz. Simulation results demonstrate that, compared to the single-modal sensing image input scheme, the multi-modal input scheme achieves superior performance in pathloss generation, with the NMSE reduced from 0.0958 to 0.0498. The improvement is attributed to the feature fusion submodule, which effectively integrates physical environment features from both RGB images and depth maps. By capturing richer environmental information, the model is better equipped to learn the complex mapping relationship between the physical environment and the electromagnetic space. Moreover, under various conditions, the proposed LLM4PG achieves an NMSE of 0.0454 in pathloss map generation, outperforming the GAN-based model baseline by more than 2.90 dB. Fig.~\ref{fs-generation} illustrates the visualized pathloss map generation results of the proposed LLM4PG compared to those of the GAN-based model baseline. Simulation results indicate that the proposed LLM4PG outperforms the GAN-based model baseline by producing pathloss maps with clearer building boundaries and more accurate pathloss values, thereby exhibiting enhanced pathloss generation performance.

\begin{table*}[!t]
\renewcommand\arraystretch{1.5}
\caption{Full-Sample Generation Performance of LLM4PG}
\centering
\begin{tabular}{|c|c|c|c|c|c|}
\hline
\textbf{Scenario} & \textbf{Flight altitude} & \textbf{Frequency band} & \textbf{Input data} &\textbf{Pathloss generation model} &\textbf{Pathloss generation NMSE} \\ 
\hline
\multicolumn{1}{|c|}{\multirow{8}{*}{\makecell[c]{Urban crossroad}}} & \multicolumn{1}{c|}{\multirow{4}{*}{50 m}} & \multicolumn{1}{c|}{\multirow{4}{*}{\makecell[c]{28 GHz}}} & \multicolumn{1}{c|}{\multirow{1}{*}{\makecell[c]{RGB image}}} & \multicolumn{1}{c|}{\multirow{1}{*}{\makecell[c]{LLM4PG}}} & {0.0564} \\ 
\cline{4-6}

{} & {} & {} & {Depth map} & {LLM4PG} & {0.0958} \\
\cline{4-6}

{} & {} & {} & \multicolumn{1}{c|}{\multirow{2}{*}{RGB-D image}} & {LLM4PG} & {\textbf{0.0498}}\\
\cline{5-6}
{} & {} & {} & {} & {GAN} & {0.1120}\\
\cline{2-6}

    \multicolumn{1}{|c|}{}& \multicolumn{1}{c|}{\multirow{4}{*}{\makecell[c]{70 m}}} & \multicolumn{1}{c|}{\multirow{2}{*}{\makecell[c]{28 GHz}}} &  \multicolumn{1}{c|}{\multirow{2}{*}{\makecell[c]{RGB-D image}}} & \multicolumn{1}{c|}{\multirow{1}{*}{\makecell[c]{LLM4PG}}} & {\textbf{0.0480}} \\
    \cline{5-6} 
    {} & {} & {} & {} & {GAN} & {0.1256} \\
    \cline{3-6}
    {} & {} & \multicolumn{1}{c|}{\multirow{2}{*}{\makecell[c]{1.6 GHz}}} & \multicolumn{1}{c|}{\multirow{2}{*}{\makecell[c]{RGB-D image}}} & {LLM4PG} & {\textbf{0.0454}} \\
    \cline{5-6}
    {} & {} & {} & {} & {GAN} & {0.0886} \\
    \hline
\multicolumn{1}{|c|}{\multirow{4}{*}{\makecell[c]{Urban wide lane}}} & \multicolumn{1}{c|}{\multirow{4}{*}{200 m}} & \multicolumn{1}{c|}{\multirow{2}{*}{\makecell[c]{28 GHz}}} & \multicolumn{1}{c|}{\multirow{2}{*}{\makecell[c]{RGB-D image}}} & \multicolumn{1}{c|}{\multirow{1}{*}{\makecell[c]{LLM4PG}}} & {\textbf{0.1877}} \\ 
    \cline{5-6} 
    
    {} & {} & {} & {} & {GAN} & {0.5694}  \\
    \cline{3-6}
    {} & {} & \multicolumn{1}{c|}{\multirow{2}{*}{\makecell[c]{1.6 GHz}}} & \multicolumn{1}{c|}{\multirow{2}{*}{\makecell[c]{RGB-D image}}} & {LLM4PG} & {\textbf{0.2014}} \\
    \cline{5-6}
    {} & {} & {} & {} & {GAN} & {0.5987} \\
    
\hline

\end{tabular}
\label{full-samples}
\end{table*}

\subsection{Pathloss Generation Performance of LLM4PG \textbf{Across Different Flight Altitudes}}

\begin{figure}[!t]
\centering
\includegraphics[width=0.4\textwidth]{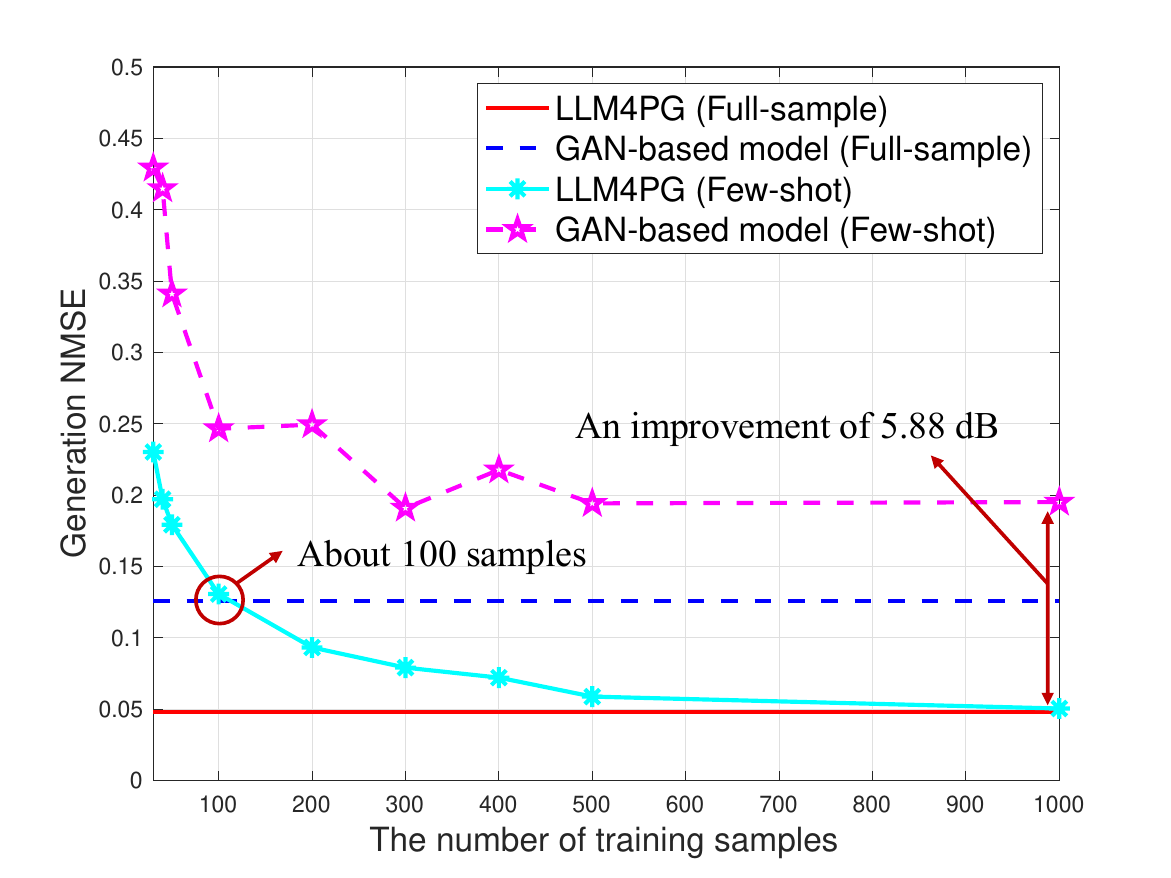}
\caption{Generalization performance across different flight altitudes from 50 m to 70 m.}
\label{fanhua-height-50-70}
\end{figure}

\begin{figure}[!t]
\centering
\includegraphics[width=0.4\textwidth]{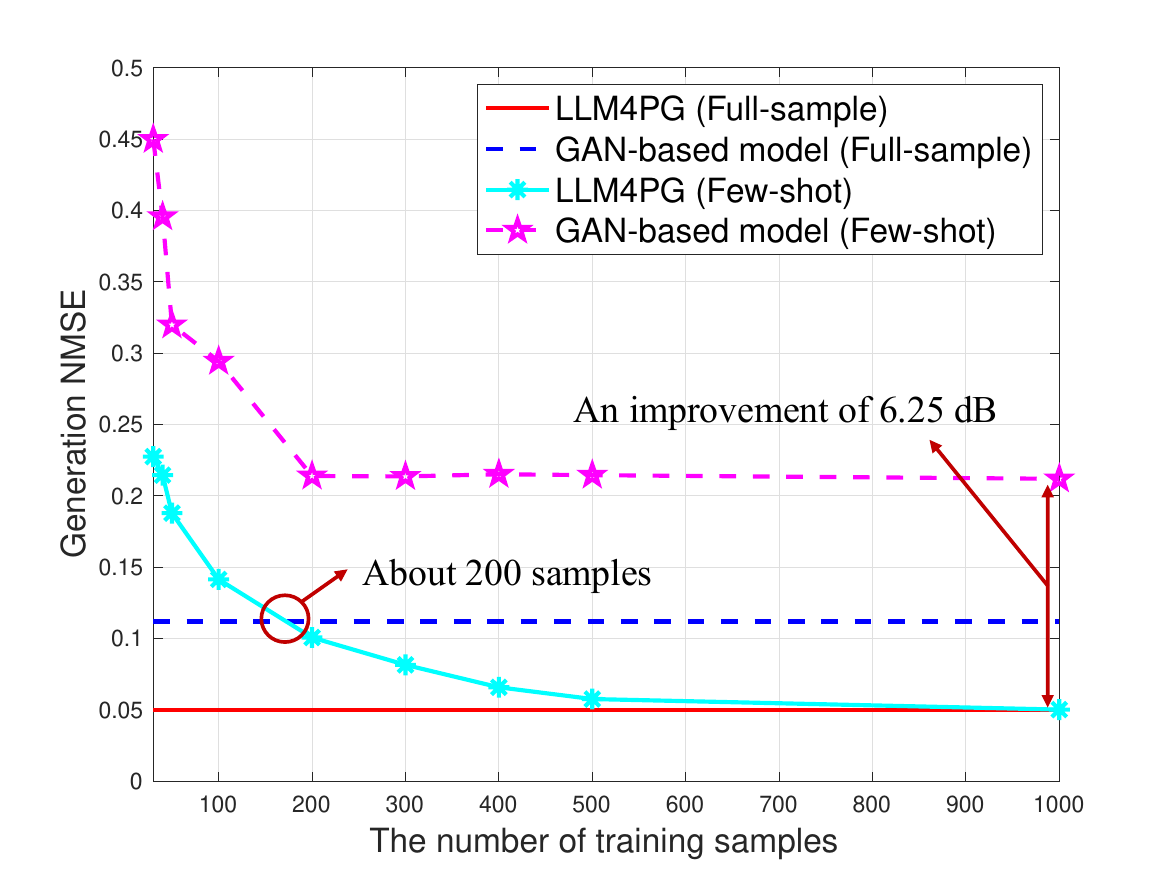}
\caption{Generalization performance across different flight altitudes from 70 m to 50 m.}
\label{fanhua-height-70-50}
\end{figure}

The transfer learning performance of the proposed LLM4PG across different flight altitudes, i.e., 50 m and 70 m, is evaluated in this subsection. In the process of transfer learning, a few-shot learning approach is employed, where the model is first trained on a dataset collected at one flight altitude, and then further trained and tested using a small number of samples from the dataset collected at the other flight altitude. This approach demonstrates the transfer generalization capability of the proposed LLM4PG across different flight altitudes, thereby enabling few-shot transfer learning for pathloss generation at varying flight altitudes. As shown in Fig.~\ref{fanhua-height-50-70} and Fig.~\ref{fanhua-height-70-50}, the generalization performance across different flight altitudes of the proposed LLM4PG is compared with that of the GAN-based model baseline at the 28 GHz frequency band in the urban crossroad scenario. Specifically, Fig.~\ref{fanhua-height-50-70} illustrates the generalization performance of LLM4PG and GAN-based baseline from 50 m to 70 m, where the model is trained on the dataset collected at 50 m flight altitude and subsequently fine-tuned on a small number of samples from the 70 m flight altitude dataset. Similarly, Fig.~\ref{fanhua-height-70-50} illustrates the generalization performance of LLM4PG and the GAN-based baseline from 70 m to 50 m. In the cross-flight altitude generalization, the proposed LLM4PG outperforms the GAN-based baseline by at least 5.88 dB when fine-tuned with 1,000 samples. Simulation results show that the proposed LLM4PG can achieve comparable performance to the GAN-based model baseline trained on the full dataset by using only a few samples for fine-tuning on the new dataset. Moreover, it exhibits superior generalization capability compared to the baseline. Note that the generalization performance from 50 m to 70 m is better than that from 70 m to 50 m, as the proposed LLM4PG reaches the full-sample performance of the GAN-based baseline using only 100 fine-tuning samples in the former case, whereas the latter requires approximately 200 samples to achieve comparable accuracy. This is because, at the lower altitude of 50 m, the field of view available to the UAV is more restricted, making it more challenging to predict pathloss in shadowed areas near buildings. As a result, training the model on the 50 m dataset enables it to learn more robust pathloss generation capabilities, which in turn facilitates better generalization to the 70 m scenario. These results demonstrate that LLM4PG maintains high generation accuracy when generalizing across different UAV flight altitudes, ensuring the availability of massive-scale and high-quality pathloss data under diverse operational flight altitudes. This flight altitude-diverse pathloss data foundation is critical for 6G AI-native communication systems, enabling altitude-adaptive communication resource allocation and robust model training without the need for substantial data collection at every flight altitude.

\subsection{Pathloss Map Generation Performance of LLM4PG \textbf{Across Different Scenarios}}

\begin{figure}[!t]
\centering
\includegraphics[width=0.4\textwidth]{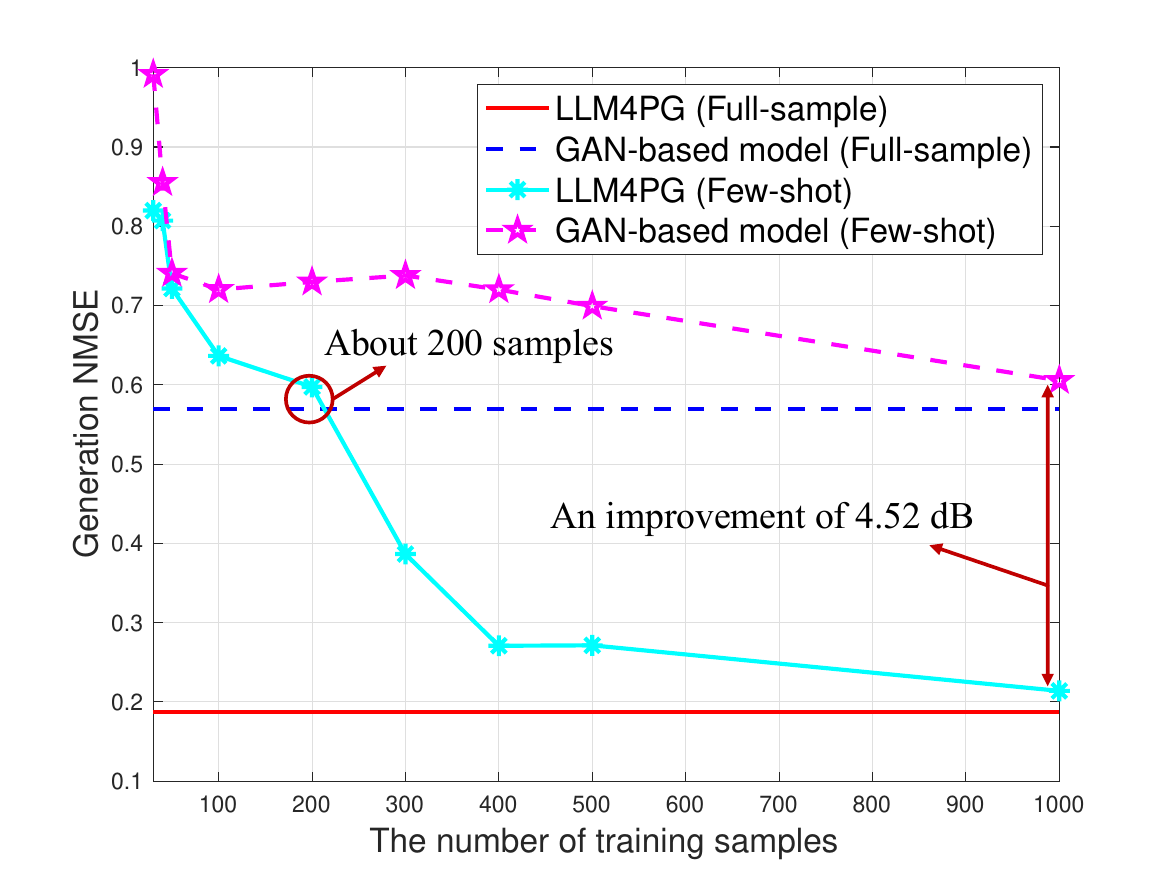}
\caption{Generalization performance across different scenarios from crossroad to wide lane.}
\label{sz-ck}
\end{figure}

\begin{figure}[!t]
\centering
\includegraphics[width=0.4\textwidth]{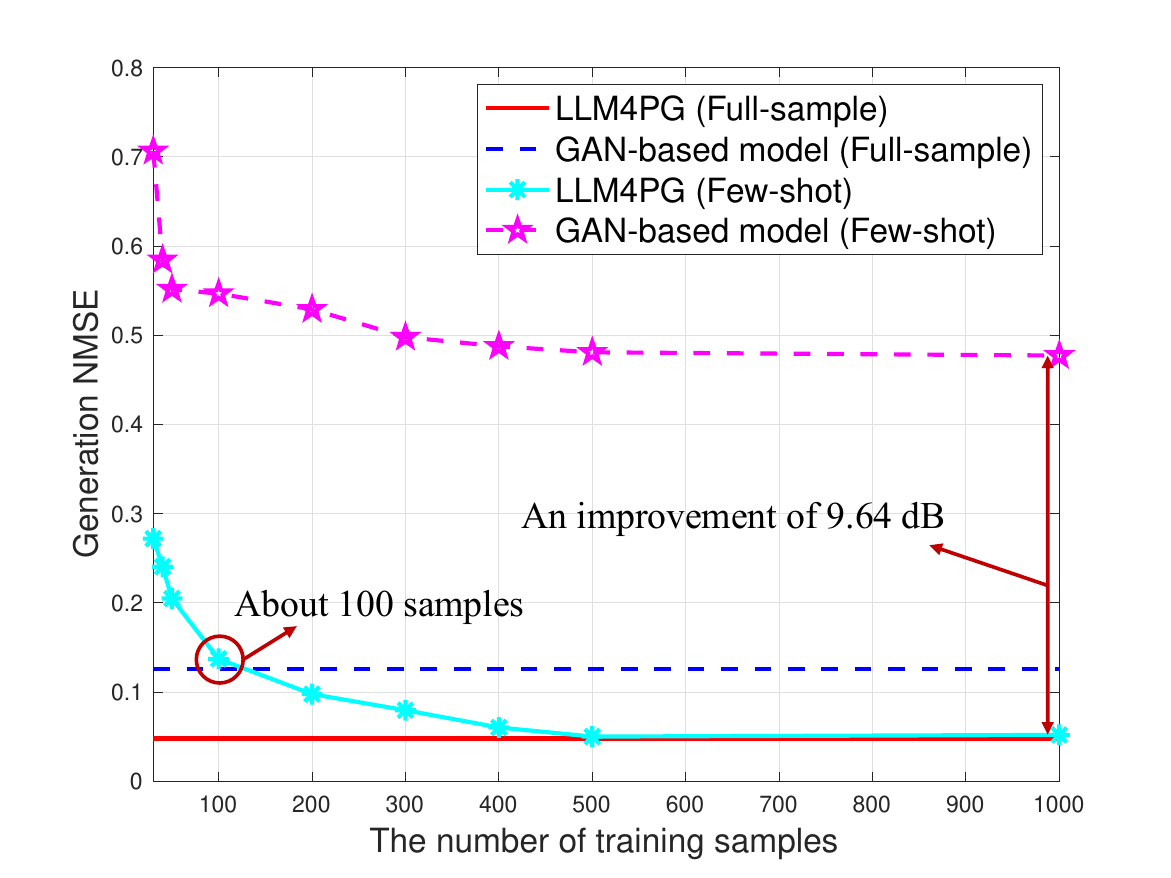}
\caption{Generalization performance across different scenarios from wide lane to crossroad.}
\label{ck-sz}
\end{figure}

The transfer learning performance of the proposed LLM4PG across different scenarios, i.e., urban crossroad scenario and urban wide lane scenario, is evaluated in this subsection. Compared to the crossroad scenario, the buildings in the wide lane scenario are taller, denser, and more complexly distributed. Similar to the cross-flight altitude generalization in Section IV-C, the cross-scenario generalization means that the model is first trained on one scenario and then undergoes few-shot learning and testing on the other scenario to evaluate its generalization performance across different scenarios. As shown in Fig.~\ref{sz-ck} and Fig.~\ref{ck-sz}, the generalization performance across different scenarios of the proposed LLM4PG and is compared with that of the GAN-based model baseline at the 28 GHz frequency band. Specifically, Fig.~\ref{sz-ck} shows the generalization performance of LLM4PG and GAN-based baseline from the crossroad scenario to the wide lane scenario, where the model is trained on the crossroad scenario and then fine-tuned on a small number of samples from the wide lane scenario. Fig.~\ref{ck-sz} illustrates the generalization performance from the wide lane scenario to the crossroad scenario. In the cross-scenario generalization, the proposed LLM4PG outperforms the GAN-based baseline by at least 4.52 dB when fine-tuned with 1,000 samples. Simulation results demonstrate that the proposed LLM4PG exhibits superior generalization across different scenarios compared to the baseline, achieving the performance of the conventional deep learning model trained with full samples through fine-tuning with only a few samples. Furthermore, as shown in Fig.~\ref{sz-ck} and Fig.~\ref{ck-sz}, generalizing from the wide lane scenario to the crossroad scenario is easier, as the proposed LLM4PG achieves the performance of the conventional deep learning model trained with full samples using only 100 fine-tuning samples, whereas the inverse transfer requires approximately 200 samples to achieve comparable accuracy. This is because the wide lane scenario is more complex, with taller buildings and more intricate spatial arrangements, leading to a more complex pathloss distribution. The proposed LLM4PG trained on the wide lane scenario can learn richer and more transferable features in physical environment and electromagnetic space. Therefore, when generalizing to the simpler crossroad scenario, the proposed LLM4PG can more easily explore the mapping relationship between RGB-D images in physical environment and pathloss in electromagnetic space. The ability of LLM4PG to generalize across distinct U2G scenarios ensures that massive-scale and high-quality pathloss maps can be generated for a variety of U2G scenarios. Such scenario-diverse data forms the essential data foundation for 6G AI-native communication systems, supporting cross-scenario model adaptation, cooperative sensing, and intelligent network planning without scenario-specific retraining.

\subsection{Pathloss Map Generation Performance of LLM4PG \textbf{Across Different Frequency Bands}}

\begin{figure}[!t]
\centering
\includegraphics[width=0.4\textwidth]{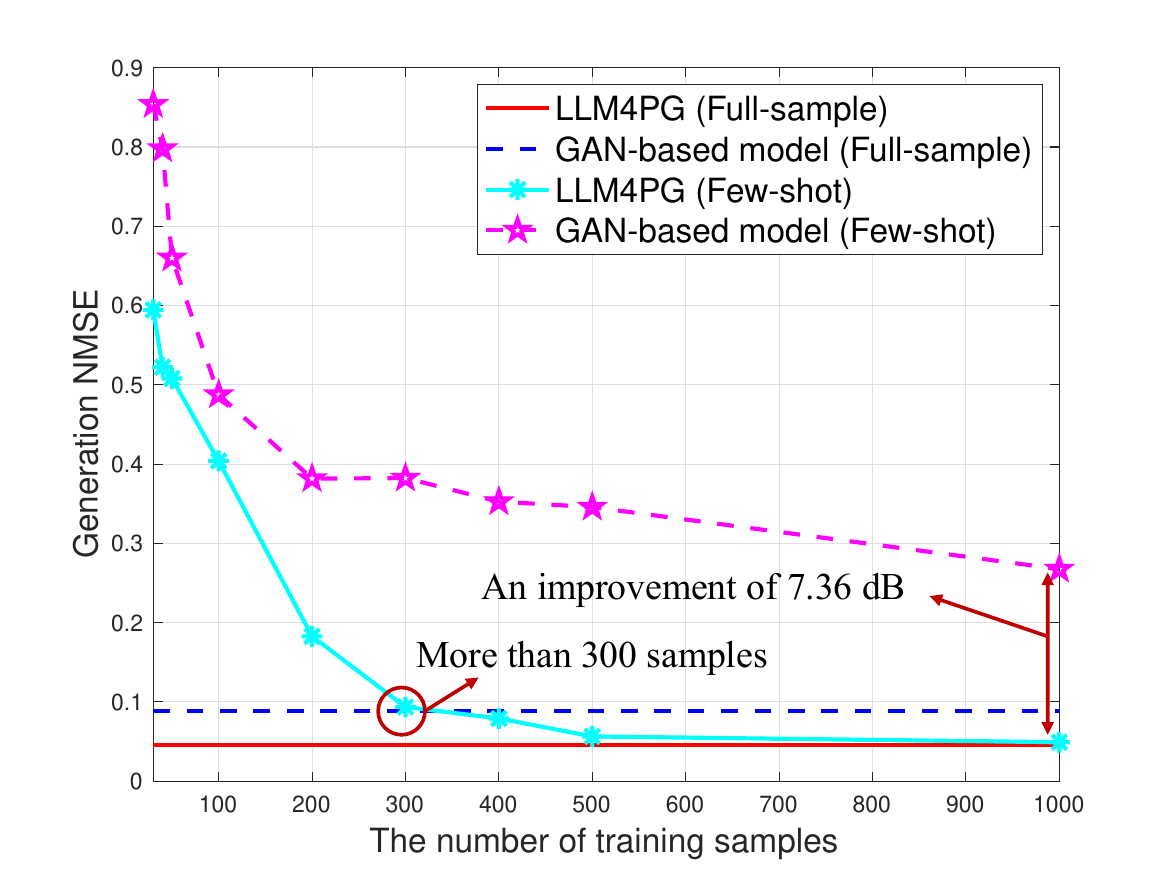}
\caption{Generalization performance across different frequency bands from 28 GHz to 1.6 GHz in urban crossroad scenario.}
\label{28-1.6-sz}
\end{figure}

\begin{figure}[!t]
\centering
\includegraphics[width=0.4\textwidth]{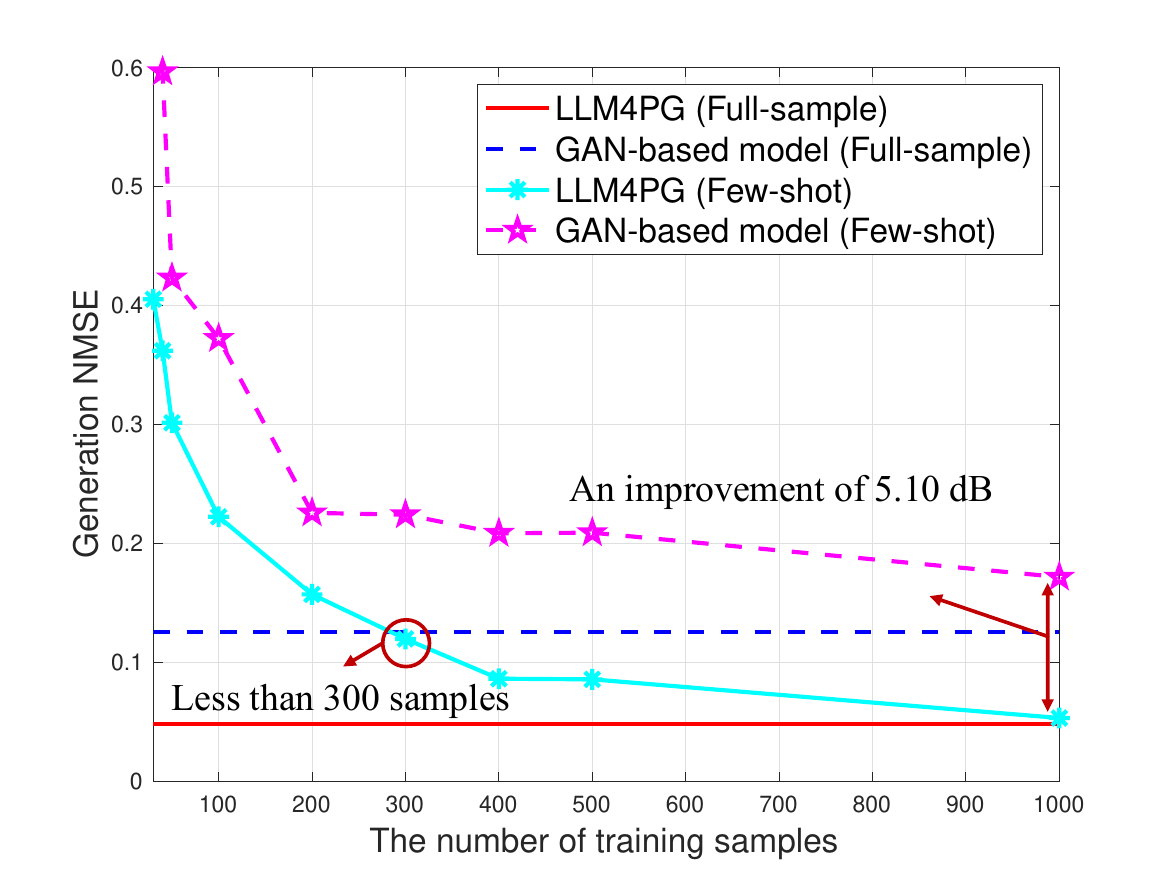}
\caption{Generalization performance across different frequency bands from 1.6 GHz to 28 GHz in urban crossroad scenario.}
\label{1.6-28-sz}
\end{figure}

\begin{figure}[!t]
\centering
\includegraphics[width=0.4\textwidth]{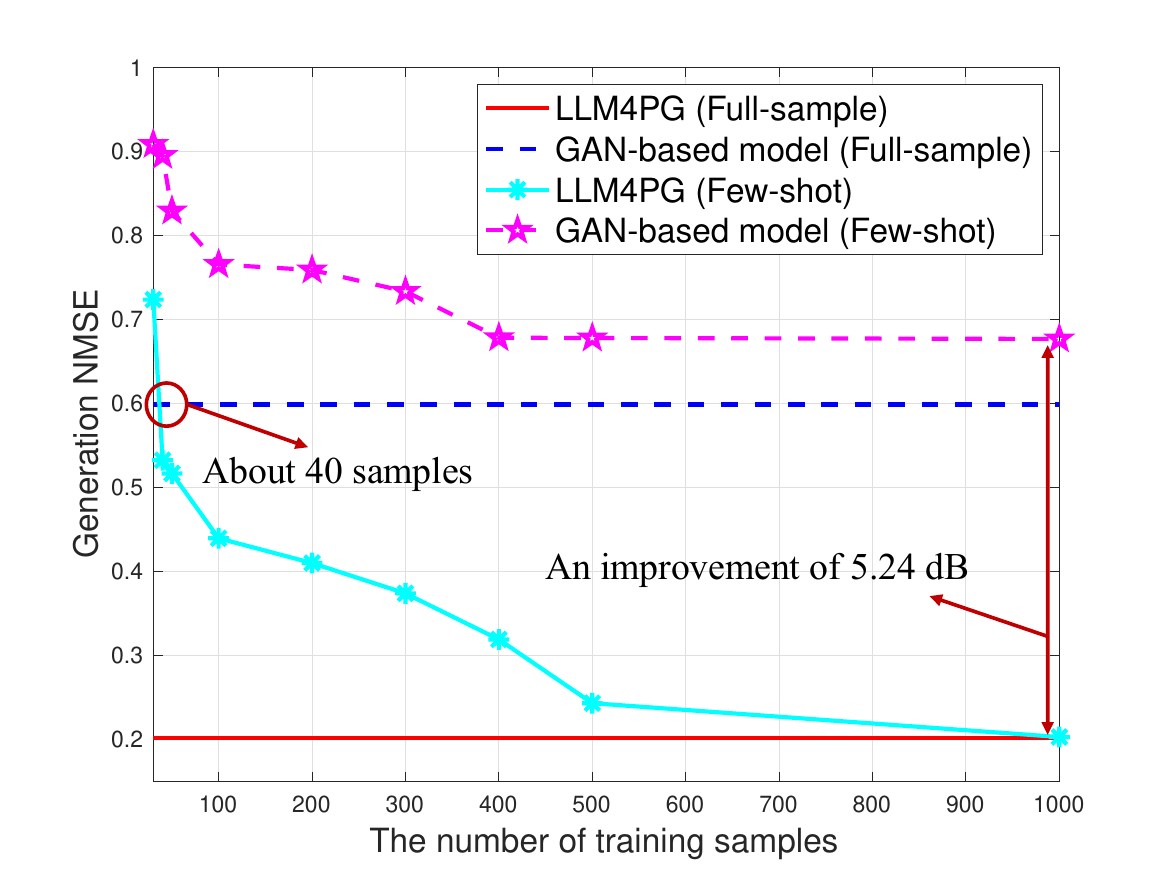}
\caption{Generalization performance across different frequency bands from 28 GHz to 1.6 GHz in urban wide lane scenario.}
\label{28-1.6-ck}
\end{figure}

The transfer learning performance of the proposed LLM4PG across different frequency bands, i.e., 28 GHz and 1.6 GHz, is evaluated in this subsection. Compared to the 28 GHz frequency band, the diffraction effect of signals in the 1.6 GHz frequency band is more pronounced, resulting in a more complex distribution of pathloss. This is particularly evident in areas where buildings obstruct the signal, making pathloss generation more challenging. Similar to flight altitude and scenario generalization, the frequency band generalization means that the model is first trained on one frequency band and undergoes few-shot learning and testing on the other frequency band to evaluate its generalization performance across different frequency bands. Fig.~\ref{28-1.6-sz} and Fig.~\ref{1.6-28-sz} respectively illustrate the generalization performance of the proposed LLM4PG when transferring from 28 GHz to 1.6 GHz and from 1.6 GHz to 28 GHz in the urban crossroad scenario at the flight altitude of 70 m. Fig.~\ref{28-1.6-ck} shows the generalization performance from 28 GHz to 1.6 GHz in the urban wide lane scenario, in comparison with the GAN-based model baseline. In the cross-frequency band generalization, the proposed LLM4PG outperforms the GAN-based baseline by at least 5.10 dB when fine-tuned with 1,000 samples. Simulation results show that the proposed LLM4PG exhibits superior generalization across different frequency bands compared to the baseline, achieving the performance of the conventional deep learning model trained with full samples through fine-tuning with only a few samples. Furthermore, Fig.~\ref{28-1.6-sz} and Fig.~\ref{1.6-28-sz} show that generalizing from 1.6 GHz to 28 GHz is easier compared to generalizing from 28 GHz to 1.6 GHz, as the proposed LLM4PG achieves the full-sample performance of the GAN-based baseline using fewer than 300 fine-tuning samples, whereas generalizing from 28 GHz to 1.6 GHz requires more than 300 samples to reach comparable accuracy. This is because the pathloss distribution in electromagnetic space at 1.6 GHz is more complex, enabling the model to learn more transferable features in electromagnetic space. As a result, the model demonstrates strong cross-modal pathloss generation capabilities when generalizing to higher frequency bands. Moreover, Fig.~\ref{28-1.6-sz} and Fig.~\ref{28-1.6-ck} show that under the certain frequency band generalization conditions, generalization in the complex scenario, i.e., the wide lane scenario, is more challenging compared to the simple scenario, i.e., the urban crossroad scenario. Specifically, in the wide lane scenario, the proposed LLM4PG achieves the full-sample performance of the GAN-based baseline with only 40 fine-tuning samples. This is because the spatial arrangement of buildings and pathloss distribution are more intricate in the wide lane scenario, making it harder to explore the mapping relationship between physical environment and electromagnetic space. Therefore, even when performing the certain frequency band generalization, i.e., from 28 GHz to 1.6 GHz, there are appropriate differences in different scenarios.  The demonstrated cross-frequency generalization enables LLM4PG to generate pathloss data at new frequency bands with minimal data requirements, significantly reducing measurement costs. This capability ensures that the 6G AI-native communication systems can access frequency-diverse and high-quality pathloss data, which are indispensable for spectrum-aware model training and dynamic spectrum management.

\subsection{Network Storage and Inference Cost}

The model complexity and time costs for both training and inference are key indicators of storage and computational overhead. Table~\ref{cost} presents a comparison between LLM4PG and the GAN-based model baseline in terms of parameters, training time, and inference time, using samples collected from the urban crossroad scenario. The proposed LLM4PG contains 83.64 M learnable parameters and 136.41 M total parameters, whereas the GAN-based baseline model contains 48.39 M learnable parameters and the same number of total parameters. It is evident that LLM4PG updates only a limited subset of its parameters during training, significantly reducing the computational burden. Moreover, its inference time remains on par with that of the baseline model.

\begin{table}[h!]
\renewcommand\arraystretch{1.2}
\centering
\caption{Model Parameters and Time Cost}
\begin{tabular}{|c|c|c|c|}
\hline
{} & \multicolumn{1}{c|}{\multirow{1}{*}{\makecell[c]{\textbf{Parameters}}}} & \multicolumn{1}{c|}{\multirow{1}{*}{\makecell[c]{\textbf{Training time}}}} & \multicolumn{1}{c|}{\multirow{1}{*}{\makecell[c]{\textbf{Inference time}}}} \\ 
& \multicolumn{1}{c|}{\multirow{1}{*}{\makecell[c]{\textbf{(M)}}}} & \textbf{(ms)} & \textbf{(ms)} \\   \hline
LLM4PG & 83.64/136.41 & 9.27 & 5.32 \\ \hline
GAN & 48.39/48.39 & 14.42 & 4.27 \\ \hline

\end{tabular}
\label{cost}
\end{table}

\section{Conclusion}
This paper has proposed a novel LLM-based pathloss map generation model, named LLM4PG, which serves as an general and effective tool for generating massive-scale and high-quality pathloss data for the 6G AI-native communication systems. A new synthetic intelligent multi-modal
sensing-communication dataset for SoM in U2G scenarios, named SynthSoM-U2G, has been constructed, including 5,660 RGB images, 5,660 depth maps, and 9,490 pathloss maps in different scenarios, frequency bands, and flight altitudes. Based on the constructed SynthSoM-U2G dataset, the proposed LLM4PG has achieved accurate pathloss generation and has demonstrated strong generalization, providing the capability to generate massive-scale and high-quality pathloss data with diversity across scenarios, frequency bands, and flight altitudes, which form the critical data foundation for the 6G AI-native communication systems. Simulation results have shown that the proposed LLM4PG has achieved accurate pathloss generation with an NMSE of 0.0454, outperforming the conventional deep learning AIGC GAN-based model by more than 2.90 dB in full-sample pathloss generation. Furthermore, simulation results have indicated that the proposed LLM4PG has demonstrated outstanding few-shot generalization capabilities with an NMSE of 0.0492, including cross-scenario, cross-frequency band, and cross-flight altitude generalization, exceeding the performance of the conventional deep learning AIGC GAN-based model by more than 4.52 dB. In the few-shot generalization, the proposed LLM4PG has achieved the full-sample performance of the conventional deep learning AIGC model-based approach using no more than 400 samples in few-shot generalization. Moreover, simulation results have shown that generalizing from more complex and challenging conditions to simpler ones, such as from complex to simple scenarios, from 1.6 GHz to 28 GHz, and from 50 m to 70 m, is generally easier and more effective than the reverse.


%

\newpage

 





\end{document}